\documentclass[12pt]{article}
\usepackage{amssymb,amsfonts}
\usepackage{graphics,psboxit,amsmath}
\usepackage{subfigure}
\usepackage{graphicx}
\usepackage{verbatim}
\usepackage{color}
\usepackage{hyperref}
\hypersetup{colorlinks}

\definecolor{darkred}{rgb}{1,0,0}
\definecolor{darkgreen}{rgb}{0,0.5,0}
\definecolor{darkblue}{rgb}{0,0,1}
\definecolor{orange}{rgb}{1,0.5,0}
\definecolor{green}{rgb}{0,1,0}
\definecolor{purple}{rgb}{.5,0,1}

\hypersetup{ colorlinks,
linkcolor=darkred,
filecolor=darkgreen,
urlcolor=darkblue,
citecolor=darkblue,
linktocpage=true }

\definecolor{markcolor}{rgb}{.25,0,1}

\definecolor{markcolor2}{rgb}{1,0,0}

\definecolor{markcolor3}{rgb}{0,1,0}

\def\hybrid{\topmargin -0pt    \oddsidemargin 0.05in %%%%%%%%%%%%%% Archive-30pt
        \headheight 0pt \headsep 0pt
        \textwidth 16.0cm      % A4 paper
        \textheight 22,0cm       % A4 paper
        \marginparwidth .875in
        \parskip 5pt plus 1pt   \jot = 1.5ex}

\hybrid

\catcode`\@=11

\def\marginnote#1{}
\newcount\hour
\newcount\minute
\newtoks\amorpm
\hour=\time\divide\hour by60 \minute=\time{\multiply\hour by60
\global\advance\minute by-\hour}
\edef\standardtime{{\ifnum\hour<12 \global\amorpm={am}%
        \else\global\amorpm={pm}\advance\hour by-12 \fi
        \ifnum\hour=0 \hour=12 \fi
        \number\hour:\ifnum\minute<10 0\fi\number\minute\the\amorpm}}
\edef\militarytime{\number\hour:\ifnum\minute<10 0\fi\number\minute}
\def\draftlabel#1{{\@bsphack\if@filesw {\let\thepage\relax
   \xdef\@gtempa{\write\@auxout{\string
      \newlabel{#1}{{\@currentlabel}{\thepage}}}}}\@gtempa
   \if@nobreak \ifvmode\nobreak\fi\fi\fi\@esphack}
        \gdef\@eqnlabel{#1}}
\def\@eqnlabel{}
\def\@vacuum{}
\def\draftmarginnote#1{\marginpar{\raggedright\scriptsize\tt#1}}

\def\draft{\oddsidemargin -.5truein
        \def\@oddfoot{\sl preliminary draft \hfil
        \rm\thepage\hfil\sl\today\quad\militarytime}
        \let\@evenfoot\@oddfoot \overfullrule 3pt
        \let\label=\draftlabel
        \let\marginnote=\draftmarginnote
   \def\@eqnnum{(\theequation)\rlap{\kern\marginparsep\tt\@eqnlabel}%
\global\let\@eqnlabel\@vacuum}  }

\def\draft2{
        \def\@oddfoot{\sl preliminary draft \hfil
        \rm\thepage\hfil\sl\today\quad\militarytime}
        \let\@evenfoot\@oddfoot \overfullrule 3pt
        \let\label=\draftlabel
        \let\marginnote=\draftmarginnote
   \def\@eqnnum{(\theequation)\rlap{\kern\marginparsep\tt\@eqnlabel}%
\global\let\@eqnlabel\@vacuum}  }

\def\preprint{\twocolumn\sloppy\flushbottom\parindent 2em
        \leftmargini 2em\leftmarginv .5em\leftmarginvi .5em
        \oddsidemargin -.5in    \evensidemargin -.5in
        \columnsep .4in \footheight 0pt
        \textwidth 10.in        \topmargin  -.4in
        \headheight 12pt \topskip .4in
        \textheight 6.9in \footskip 0pt
        \def\@oddhead{\thepage\hfil\addtocounter{page}{1}\thepage}
        \let\@evenhead\@oddhead \def\@oddfoot{} \def\@evenfoot{} }

\def\numberbysection{\@addtoreset{equation}{section}
        \def\theequation{\thesection.\arabic{equation}}}

\def\underline#1{\relax\ifmmode\@@underline#1\else
        $\@@underline{\hbox{#1}}$\relax\fi}

\def\titlepage{\@restonecolfalse\if@twocolumn\@restonecoltrue\onecolumn
     \else \newpage \fi \thispagestyle{empty}\c@page\z@
        \def\thefootnote{\fnsymbol{footnote}} }

\def\endtitlepage{\if@restonecol\twocolumn \else \newpage \fi
        \def\thefootnote{\arabic{footnote}}
        \setcounter{footnote}{0}}  %\c@footnote\z@ }

\catcode`@=12 \relax

\def\figcap{\section*{Figure Captions\markboth
        {FIGURECAPTIONS}{FIGURECAPTIONS}}\list
        {Figure \arabic{enumi}:\hfill}{\settowidth\labelwidth{Figure
999:}
        \leftmargin\labelwidth
        \advance\leftmargin\labelsep\usecounter{enumi}}}
 \relax
\def\tablecap{\section*{Table Captions\markboth
        {TABLECAPTIONS}{TABLECAPTIONS}}\list
        {Table \arabic{enumi}:\hfill}{\settowidth\labelwidth{Table
999:}
        \leftmargin\labelwidth
        \advance\leftmargin\labelsep\usecounter{enumi}}}
 \relax
\def\reflist{\section*{References\markboth
        {REFLIST}{REFLIST}}\list
        {[\arabic{enumi}]\hfill}{\settowidth\labelwidth{[999]}
        \leftmargin\labelwidth
        \advance\leftmargin\labelsep\usecounter{enumi}}}
 \relax
\makeatletter
\newcounter{pubctr}
\def\publist{\@ifnextchar[{\@publist}{\@@publist}}
\def\@publist[#1]{\list
        {[\arabic{pubctr}]\hfill}{\settowidth\labelwidth{[999]}
        \leftmargin\labelwidth
        \advance\leftmargin\labelsep
        \@nmbrlisttrue\def\@listctr{pubctr}
        \setcounter{pubctr}{#1}\addtocounter{pubctr}{-1}}}
\def\@@publist{\list
        {[\arabic{pubctr}]\hfill}{\settowidth\labelwidth{[999]}
        \leftmargin\labelwidth
        \advance\leftmargin\labelsep
        \@nmbrlisttrue\def\@listctr{pubctr}}}
 \relax
\makeatother

\def\be{\begin{equation}}
\def\ee{\end{equation}}
\def\ba{\begin{eqnarray}}
\def\ea{\end{eqnarray}}

\def\l{\lambda}

\def\s{\sigma}

\def\cN{{\cal N}}

\def\no{\noindent}

\def\IR{\relax{\rm I\kern-.18em R}}

\def\bse{\begin{small}\begin{equation*}}
\def\ese{\end{equation*}\end{small}}

\begin{document}
%\draft2

%\renewcommand{\theequation}{\arabic{equation}}
%\renewcommand{\theequation}{\thesection.\arabic{equation}}

\renewcommand{\theequation}{\thesection.\arabic{equation}}
\csname @addtoreset\endcsname{equation}{section}

\newcommand{\eqn}[1]{(\ref{#1})}

\begin{titlepage}
\begin{center}
\strut\hfill
\vskip 1.3cm

%\hfill  [hep-th]\\

\vskip .5in

{\Large \bf A note on $\mathfrak{gl}_{\cN}$ type-I integrable defects}

\vskip 0.5in

{\large \bf Anastasia Doikou}\phantom{x}
\vskip 0.02in
{\footnotesize Department of Engineering Sciences, University of Patras,\\
GR-26500 Patras, Greece}
\\[2mm]
\noindent

\vskip .1cm

%\vskip -.15in

{\footnotesize {\tt E-mail: adoikou@upatras.gr}}\\

\end{center}

\vskip 1.0in

\centerline{\bf Abstract}
Type-I quantum defects are considered in the context of the $\mathfrak{gl}_{\cN}$ spin chain. The type-I defects are associated to the generalized harmonic oscillator algebra, and the chosen defect matrix is the one of the vector non-linear Schr\"{o}dinger (NLS) model. The transmission matrices relevant to this particular type of defects are computed via the Bethe ansatz methodology.

\no

\vfill

\end{titlepage}
\vfill \eject

\tableofcontents

\section{Introduction}
Integrable quantum defects in spin chain systems are well understood objects, and are adeptly described through the quantum inverse scattering formulation \cite{YBE}. And although many studies have been devoted to this issue at the quantum level \cite{AndreiJohannesson}--\cite{annecydef2}, the computation of related physical quantities, such as the transmission amplitudes and transmission matrices had not been attacked in this particular frame until very recently.
The transmission matrices physically describe the interaction between the particle-like excitations displayed by the integrable  model under study and the defect.
In a series of recent papers \cite{doikou-karaiskos-transmission, doikou-transmission, doikou-transmission2} the Bethe ansatz formulation was exploited for the computation of exact transmission matrices in a variety of spin chain systems, and for distinct types of defects. It is however worth noting that investigations of transmission matrices in integrable systems using the Fateev-Zamolodchikov algebra have been known for quite some time now \cite{delmusi, konle}. Similarly, at the classical level there is a wealth of relevant studies, where the distinct types of integrable defects are treated with the use of various available techniques \cite{cozanls}--\cite{doikou-karaiskos-LL}.

It will be instructive to introduce the two distinct types of defects that have been treated up to date at both quantum and classical level; these are known as type-I, and type-II defects.
Type-I defects are associated to the generic quantum oscillator algebras as well as their $q$ deformations, the type-II defects on the other hand are related to representations of the $\mathfrak{gl}_{\cN}, \ {\mathfrak U}_q(\mathfrak{gl}_{\cN})$ algebras. Type-I transmission matrices are the most well studied ones dating back to the first investigations on the subject (see e.g. \cite{konle}). Type-II transmission matrices were derived in \cite{corrigan} for the first time in the sine-Gordon context, whereas analogous results were obtained in the spin chain framework in \cite{doikou-karaiskos-transmission, doikou-transmission, doikou-transmission2}. More precisely, in the first paper of the series \cite{doikou-karaiskos-transmission, doikou-transmission, doikou-transmission2}, the Bethe ansatz frame for the study of transmission amplitudes was set. The examples worked out in \cite{doikou-karaiskos-transmission} were the XXX and XXZ spin chains in the presence of type-II defects. In \cite{doikou-transmission} type-II transmission matrices were identified for both $\mathfrak{gl}_{\cN}$, and  ${\mathfrak U}_q(\mathfrak{gl}_{\cN})$ algebras, while  in the most recent article \cite{doikou-transmission2} type-I transmission matrices were derived for the XXX and XXZ (critical and non-critical) spin chains. In the present investigation we basically complete our analysis in the context of $\mathfrak{gl}_{\cN}$ spin chain by introducing and studying the type-I defects. Our main aim here is to provide explicit expressions of transmission amplitudes and the respective type-I transitions matrices, exploiting primarily the Bethe ansatz methodology in the thermodynamic limit. The so-called ``crossing property'', which will be suitably formulated subsequently will be also used to further confirm our results.

It is worth pointing out that an exhaustive classification of the possible representations of the fundamental algebraic relation
\cite{YBE}
\be
R_{12}(\lambda_1 -\lambda_2)\ L_1(\lambda_1)\ L_2(\lambda_2) = L_2(\lambda_2)\ L_1(\lambda_1)\ R_{12}(\lambda_1 -\lambda_2)\, ,
\label{basicRLL}
\ee
associated to the $\mathfrak{gl}_{\cN}, \ {\mathfrak U}_q(\mathfrak{gl}_{\cN})$ $R$-matrix would provide novel transmission matrices.
This is a particularly pertinent problem, especially in the trigonometric case, however this will be left for future investigations.

\section{The $\mathfrak{gl}_{\cal N}$ spin chain}

As described in detail in earlier works (see e.g. \cite{doikou-karaiskos-transmission}) in order to construct an one dimensional discrete integrable system in the presence of a point-like defect on the $n^{th}$ site one needs to introduce a modified monodromy matrix, which reads then as \cite{YBE}
\be
T(\lambda) = R_{0N+1}(\lambda)\ R_{0N}(\lambda) \ldots  L_{0n}(\lambda-\Theta) \ldots R_{01}(\lambda)\, ,
\label{basic0}
\ee
where $R$ corresponds to the ``bulk'' theory, $L$ corresponds to the defect,
and $\Theta$ is an arbitrary constant corresponding to the ``rapidity'' of the defect.
The Lax operator satisfies the quadratic algebra (\ref{basicRLL}),
and the $R$-matrix is a solution of the Yang-Baxter equation (see e.g. \cite{YBE} and references therein).
The monodromy matrix of the theory $T(\l)$, naturally satisfies (\ref{basicRLL}), guaranteeing the
integrability of the model.

We shall focus in the present investigation in the $\mathfrak{gl}_{\cN}$ spin chain. The corresponding $R$-matrix is expressed in the familiar form \cite{yang}
\be
R(\lambda) = \lambda + i{\cal P}, ~~~~ {\cal P} = \sum_{k, l=1}^{\cN} e_{kl} \otimes e_{lk},
\ee
${\cal P}$ is the permutation operator associated to $\mathfrak{gl}_{\cN}$: ${\cal P}\ | a\rangle \otimes |b \rangle =  |b \rangle \otimes | a\rangle$, also $e_{kl}$ are $\cN \times \cN$ matrices with elements defined as: $\ (e_{kl})_{mn}= \delta_{km}\ \delta_{ln}$.

We choose to consider in the present investigation the generic $\mathfrak{gl}_{\cN}$ harmonic oscillator defect matrix,
or the so called discrete vector non-linear Schrodinger (NLS) $L$ matrix \cite{kunrag}:
\ba
&& L(\l) =  e_{11} \otimes (\l + i + i {\mathbb N}) + i \sum_{j=2}^{\cN} e_{jj} \otimes {\mathbb I} + i \sum_{j=2}^{\cN} \Big (e_{1j}\otimes a^{(j-1)} + e_{j1} \otimes a^{\dag(j-1)} \Big ),\cr
&& {\mathbb N} = \sum_{j=1}^{\cN-1} a^{(j)}\ a ^{\dag(j)}. \label{LL}
\ea

A significant piece of information  in our investigation, which will further confirm the validity of our expressions is the formulation of the ``crossing property''. In order to formulate the ``crossing property'' in the context of any $\mathfrak{gl}_{\cN}$ integrable system we need to introduce the conjugate $L$-matrix, which is defined as
\be
\hat L(\l) = V_1\ L^{t_1}(-\l - {i\cN \over 2}) V_1, ~~~~~V= \mbox{antidiag}(1, \ldots , 1), \label{cross}
\ee
and turns out to have the following explicit form
\be
\hat L(\l) = e_{\cN \cN} \otimes (-\l -{i\cN\over 2}+ i + i {\mathbb N}) + i \sum_{j=2}^{\cN} e_{\bar j \bar j} \otimes {\mathbb I} + i \sum_{j=2}^{\cN} \Big (e_{\bar j \cN}\otimes a^{(j-1)} + e_{\cN \bar j} \otimes a^{\dag(j-1)} \Big ). \label{hLL}
\ee
We also define the conjugate index as : $\bar j = \cN +1 -j$. Note that $\hat L$ also satisfies the fundamental algebraic relation
(\ref{basicRLL}). The ``crossing'' property (\ref{cross}) that connects the $L,\ \hat L$ matrices is essential as will be transparent later in the text when deriving the transmission matrix, and provides an extra validity check on the derived results. The chosen $L$-matrix is associated to the generalized harmonic oscillator algebra, which is expressed as:
\ba
&& \Big [a^{(i)},\ a^{\dag(j)} \Big ] =\delta_{ij}, \cr
&&\Big [{\mathbb N}, a^{(j)} \Big ] =- a^{(j)}, \cr
&& \Big [{\mathbb N}, a^{\dag(j)} \Big ] =a^{\dag(j)},\cr
&& \Big [a^{(i)},\ a^{(j)} \Big ] =0,  \cr
&&\Big [a^{{\dag}(i)},\ a^{{\dag}(j)} \Big ] =0.
\ea

The first step into deriving the transmission matrices is to extract the respective Bethe ansatz equations (BAE) via the algebraic Bethe ansatz methodology. To achieve this we assume the existence of local highest weight states such that:
\ba
&&a^{\dag(j)}\ |\omega \rangle_n =0,\cr
&&{\mathbb N}\ |\omega \rangle_n =0, \cr
&&e_{kl}\ |\omega\rangle_j = 0, ~~~~~k<l \cr
&& e_{kk}\ |\omega\rangle_j = |\omega\rangle_j, ~~~~~~~j\neq n.
\ea
The global reference state then is
\be
|\Omega \rangle = \otimes_{j=1}^{N+1} |\omega \rangle_j.
\ee
The BAEs' may be extracted as analyticity conditions of the spectrum within
the algebraic Bethe ansatz formulation. If we choose the $L$-matrix (\ref{LL}) as the defect matrix the BAEs' turn out to have the following form:
\ba
&& {\mathfrak e}^{+}(\lambda_i^{(k)} -\Theta )\delta_{k1} +1 - \delta_{k1} =\cr
&& -\prod_{j=1}^{M_{k-1}} e_{-1}(\lambda^{(k)}_i - \lambda^{(k-1)}_j)\
\prod_{j=1}^{M_k} e_2(\lambda_i^{(k)} - \lambda^{(k)}_j)\
\prod_{j=1}^{M_{k+1}} e_{-1}(\lambda^{(k)}_i - \lambda^{(k+1)}_j), \label{BAE1b}
\ea
whereas the BAEs' associated to the $\hat L$-matrix (\ref{hLL}) are given as
\ba
&& {\mathfrak e}^{-}(\lambda^{(k)}_i -\Theta )\delta_{k \cN-1} +1 - \delta_{k \cN -1} =\cr
&& -\prod_{j=1}^{M_{k-1}} e_{-1}(\lambda^{(k)}_i - \lambda^{(k-1)}_j)\
\prod_{j=1}^{M_k} e_2(\lambda^{(k)}_i - \lambda^{(k)}_j)\
\prod_{j=1}^{M_{k+1}} e_{-1}(\lambda^{(k)}_i - \lambda^{(k+1)}_j) \cr
&& \mbox{where} ~~~~~e_n(\l) = {\l + {in \over 2} \over \l -{in \over 2}}, ~~~~~~~{\mathfrak e}^{+}(\l) = \l + {i\over 2}, ~~~~~~{\mathfrak e}^{-}(\l) = {1\over \l - {i\over 2}}, \label{BAE2b}
\ea
here for simplicity we consider $\lambda_j^{(0)}=0$. Also, we define $M_0 =N$, $M_{\cN} =0$. Having identified the associated BAEs' we are now in the position to derive the relevant transmission amplitudes, and the corresponding
transmission matrices.

\subsection{The transmission matrices}

The main aim in this section is the exact derivation of type-I transmission matrices through the study of the Bethe ansatz equations in the thermodynamic limit. Note that in \cite{doikou-transmission} type-II defects were studied using the Bethe ansatz formulation. This is the first time to our knowledge that type-I defects are derived for any $\mathfrak{gl}_{\cN}$ integrable system. To extract the transmission amplitudes  it is sufficient to consider the state with one hole \cite{doikou-nepo-mezi1} --particle-like excitation-- in the first Fermi sea with rapidity $\tilde \l^{(1)}$. Recall that the ground state in this case consists of $\cN-1$ filled Fermi seas (see e.g. \cite{kure, doikou-nepo-sun} and references therein). As is well known a hole in the first sea corresponds to an excitation that
carries the fundamental representation of $\mathfrak{gl}_{\cN}$ (soliton), note also that the hole in the $(\cN-1)^{th}$ sea corresponds
to an excitation that carries the anti-fundamental (conjugate) representation.
The one-hole configuration in the first Fermi sea enables the computation of the associated transmission amplitudes.
From the BAEs' in the thermodynamic limit one derives the corresponding density of the state. This is obtained via the BAEs' with the standard process \cite{FT, Andrei-Destri}, and it reads as\footnote{To derive the densities in the thermodynamic limit we have used the following basic formula in the presence of ${\mathrm m}^{(k)}$ holes in the $k^{th}$ Fermi sea:
\be
{1\over N} \sum_{j=1}^{M^{(k)}}f(\lambda_j^{(k)}) \to \int_{-\infty}^{\infty}d\lambda\ f(\lambda)\ \s^{(k)}(\lambda) -
{1\over N} \sum_{i=1}^{{\mathrm m}^{(k)}} f(\tilde \lambda_i^{(k)}).
\label{basic}
\ee}
\be
\sigma^{\pm(k)}(\lambda) = \sigma_0^{(k)}(\lambda) + {1\over N} \Big ( r^{(k)}(\lambda -\tilde \lambda^{(1)}) +
r_t^{\pm(k)}(\lambda -\Theta) \Big ). \label{densk}
\ee
The plus in the densities above corresponds to (\ref{BAE1b}), while the minus corresponds to (\ref{BAE2b}).
The Fourier transforms of the quantities involved in (\ref{densk}) are given by
\ba
&& \hat \s_0^{(k)}(\omega) = {\sinh((\cN - k){\omega \over 2}) \over \sinh ({\cN\omega \over 2}) },
~~~~~~~k\in \Big \{1, 2 , \ldots, \cN-1 \Big \} \cr
&& \hat r^{(k)}(\omega) = \hat R_{k1}(\omega)\ \hat a_2(\omega) - \hat R_{k2}(\omega)\ \hat a_1(\omega), \cr
&& \hat r_t^{+(k)}(\omega)=  \hat R_{k1}(\omega)\ \hat {\mathfrak a}^+(\omega) \cr
&& \hat r_t^{-(k)}(\omega)= \hat R_{k \cN-l}(\omega)\ \hat {\mathfrak a}^-(\omega).
 \label{densr}
\ea
We also introduce the following important quantities:
\be
a_n(\lambda) = {i \over 2 \pi } {d \over d\lambda}\ln \Big (e_n(\lambda) \Big ), ~~~~
{\mathfrak a}^{\pm}(\l) = {i\over 2\pi} {d\over d\lambda} \ln\Big( {\mathfrak e}^{\pm}(\lambda) \Big)
\ee
and
\ba
&& \hat R_{jj'}(\omega) = {e^{{|\omega|\over 2}} \sinh\Big ({j_< \omega \over 2}\Big )\
\sinh \Big ((\cN - j_>){\omega \over 2}\Big )\over \sinh({\omega \over 2})\ \sinh\Big ({\cN \omega \over 2} \Big )},
~~~~~\hat a_n(\omega) = e^{-n {|\omega| \over 2}}, \cr && j_< = \mbox{min} \{j,\ j'\}, ~~~j_> = \mbox{max} \{j,\ j'\}\cr
&&\hat {\mathfrak a}^{+}(\omega) = e^{\omega \over 2} ~~~~\omega < 0, ~~~~~\hat {\mathfrak a}^{+}(\omega) =0 ~~~~~\omega>0, \cr
&&\hat {\mathfrak a}^{-}(\omega) = e^{-{\omega \over 2}} ~~~~\omega > 0, ~~~~~\hat {\mathfrak a}^{-}(\omega) =0 ~~~~~\omega<0.
\ea

We are naturally interested in the densities of the first Fermi sea, which provide the transmission amplitude that describes the interaction between the defect and the particle-like excitation (hole in the first Fermi sea). Recall also that
\be
\sigma^{(k)}_0(\lambda) = \varepsilon^{(k)}(\lambda), ~~~~\mbox{and}~~~~~~
\varepsilon^{(k)}(\lambda) = {1\over 2 \pi}{d p^{(k)}(\lambda)\over d \lambda} \, ,\label{EP}
\ee
with $\varepsilon^{(k)}$ and $p^{(k)}$ being the energy and the momentum of the hole
excitation in the $k$th sea, respectively. Moreover the densities $\sigma^{(k)}$ are defined as
\be
\sigma^{(k)}(\lambda) = {1\over N}{ d h^{(k)}(\lambda) \over d\lambda} \label{sigma}
\ee
$h^{(k)}(\lambda)$ is the so-called counting function and $h^{(k)}(\tilde \lambda^{(k)}_i) = J^{(k)}_i$, where $J^{(k)}_i$ are integer numbers.

In order to derive the hole-defect transmission amplitude,
we compare the expression providing the density of the first Fermi sea (\ref{densk}) with the so called quantization condition for a state with one particle \cite{FT, Andrei-Destri}. This condition with respect to the
hole with rapidity $\tilde \lambda^{(1)}$ is expressed then as
\be
\Big (e^{i N p^{(1)}(\tilde \lambda^{(1)})}\ {\mathfrak T}(\tilde \lambda^{(1)}, \Theta) -1 \Big )
|\tilde \lambda^{(1)}\rangle =0 \, ,
\label{QC1}
\ee
$p^{(1)}(\tilde \lambda^{(1)})$ is the momentum of the respective hole in the first Fermi sea. Moreover, ${\mathfrak T} \in \{{\mathbb T},\ \bar {\mathbb T} \}$. ${\mathbb T}$ and $\hat {\mathbb T}$ are the relevant transmission matrices, depending on the choice of the defect matrix $L$ or $\hat L$ respectively.
Comparison of the quantization condition with the state's density (\ref{densk}) immediately provides the transmission amplitudes (eigenvalues) (see also \cite{doikou-karaiskos-transmission, doikou-transmission, doikou-transmission2}).
The transmission amplitudes for the model with a single defect can be then derived as
\be
T^{\pm}(\hat \lambda)=  \exp \Big [ - \int_{-\infty}^{\infty}\  {d\omega \over \omega} e^{-i\omega \hat\lambda}
\hat r_{t}^{\pm(1)}(\omega)\Big ] \label{ST}
\ee
where we have set $\hat \lambda = \tilde \lambda_1^{(1)} - \Theta$, and the $\pm$ in $T^{\pm}$ correspond to the $\pm$ in (\ref{densk}).

Bearing also in mind the useful identity
\be
{1\over 2} \int_0^{\infty}\ {dx \over x} {e^{-{\mu x\over 2}}\over \cosh({x\over 2})} =
\ln {\Gamma({\mu +1\over 4}) \over \Gamma({\mu +3\over 4})}\, ,
\label{ident}
\ee
as well as formulas (\ref{densr}) we conclude that the transmission amplitudes are identified via (\ref{ST}), and have the
following explicit forms:
\be
T^+(\lambda) = {\Gamma(-{i\l \over \cN} + { 1 \over 2 \cN})\over
\Gamma(-{i\l \over \cN} - { 1\over 2 \cN} +1)}, ~~~~~~~T^-(\lambda) = {\Gamma({i\l \over \cN} + { 1 \over 2 \cN} +{1\over 2})\over
\Gamma({i\l \over \cN} - { 1\over 2 \cN} +{1\over 2})}. \label{T1}
\ee

Keep in mind that for purely transmitting defects the following quadratic algebra is satisfied by the transmission matrices
\cite{delmusi}
\be
{\mathbb S}_{12}(\lambda_1 -\lambda_2)\ {\mathbb T}_1(\lambda_1)\ {\mathbb T}_2(\lambda_2) =
{\mathbb T}_2(\lambda_2)\ {\mathbb T}_1(\lambda_1)\ {\mathbb S}_{12}(\lambda_1 -\lambda_2)\, .
\label{rttb}
\ee
In fact, the latter relation encodes the mathematical content associated to the transmission matrices; they are apparently representations of the quadratic algebra (\ref{rttb}).
In our case the ${\mathbb S}$-matrix is the typical $\mathfrak{gl}_{\cN}$ scattering matrix \cite{yang}:
\be
{\mathbb S}(\lambda) = {S(\lambda) \over i\l +1} \Big (i\l + {\cal P} \Big), \label{Smatrix}
\ee
where the  hole-hole scattering amplitude $S$ is given as (see e.g. \cite{doikou-nepo-sun, ogrewi}):
\be
S(\lambda) = {\Gamma({i\l\over \cN} +1 )\  \Gamma(-{i\l\over \cN} +1 -{1\over \cN})\over
\Gamma(-{i\l\over \cN} +1 )\ \Gamma({i\l\over \cN} +1 -{1\over \cN})}.
\ee
Taking the latter information into account, together with the crossing property, we conclude that the transmission matrix may be cast as
\ba
&& {\mathbb T}(\lambda ) = {T^-(\lambda) \over i\l+ {\cN \over 2} - {1\over 2} } \Big (e_{11} \otimes (i\l +1 +\bar {\mathbb N}) + \sum_{j=2}^{\cN} e_{jj} \otimes {\mathbb I} + \sum_{j=2}^{\cN}(e_{1j} \otimes a^{(j)} + e_{j1}\otimes a^{\dag(j-1)})\Big), \cr
&& \bar {\mathbb N} = \sum_{j=1}^{\cN-1}a^{(j)}a^{\dag(j)} +{\cN \over 2} - {3\over 2},  \label{TT}
\ea
whereas the``conjugate'' transmission matrix is of the expected form:
\ba
&& \bar {\mathbb T}(\lambda) = T^+(\lambda)
\Big (e_{\cN \cN}\otimes (-i\l - {\cN \over 2} +1 +\bar {\mathbb N}) + \sum_{j=2}^{\cN} e_{\bar j \bar j}\otimes {\mathbb I} + \sum_{j=2}^{\cN} (e_{\bar j \cN}\otimes a^{(j-1)} + e_{\cN\bar j}\otimes a^{\dag(j-1)} )\Big ).\cr &&\label{bTT}
\ea
Note that as in the $\mathfrak{sl}_2$ case investigated in \cite{doikou-transmission2} the ``physical'' quantity $\bar {\mathbb N}$ is shifted compared to the ``bare'' quantity ${\mathbb N}$ by the constant ${\cN \over 2} - {3\over 2}$, which is also confirmed by a direct computation via the BAEs'.

Some comments are in order here; it is clear that in general (\ref{TT}), (\ref{bTT}) satisfy the quadratic relation (\ref{rttb}) for any overall factor in front of the matrices. With the use of BAEs' we fix exactly this overall physical factor by computing explicitly the eigenvalues of the matrix. The validity of the physical overall factor may be further checked via the crossing property that should be satisfied by the transmission matrices.
More precisely, the overall factors in front of the transmission matrices above are compatible with the BAEs', and they also satisfy the crossing property as will be clear immediately below.
Indeed it is easy to verify that
\be
 \bar {\mathbb T}_{12}(\lambda)= {\cal C}\ V_1\ {\mathbb T}_{12}^{t_1}(-\lambda +{i\cN \over 2 })\ V_1 \label{crossb}
\ee
as expected from the crossing property, ${\cal C}$ is $\l$-independent arbitrary constant. Confirmation of the crossing property provides an extra validity check on our results. As already discussed in the $\mathfrak{sl}_2$ case, studied in \cite{doikou-transmission2}, the ${\mathbb T}$ and $\bar {\mathbb T}$ matrices are equivalent algebraic objects due to the crossing property (\ref{cross}), as well as the fact that they satisfy the same quadratic algebra  (\ref{rttb}). This concludes our derivation of type-I transmission matrices within the $\mathfrak{gl}_{\cN}$ spin chain.

\section{Conclusions}
In summary, we have been able to explicitly identify the type-I transmission matrices within the $\mathfrak{gl}_{\cN}$ spin chain frame exploiting mainly the algebraic Bethe ansatz framework as well as the crossing property for the generalized oscillator algebra incorporated in the defect matrices $L,\ \hat L$. We have restricted our attention here in the isotropic case, however similar results can be extracted for the trigonometric case, provided that a $q$-deformation of the vector non-linear Schr\"{o}dinger model is first derived. This information is still missing to our knowledge, hence this remains an open problem.

In general, in a series of recent papers \cite{doikou-karaiskos-transmission, doikou-transmission, doikou-transmission2} both known types of defects I and II have been treated with the use of the Bethe ansatz framework, and the relevant physical transmission amplitudes were derived in both isotropic and trigonometric cases, with the exception of the study of type-I defects in the higher rank trigonometric case, as already pointed out. The situation therefore within the spin chain frame is more or less well understood and controlled. Clearly the findings of the present investigation may be mapped in a straightforward manner to integrable quantum field theories, such as the Gross-Neveu model or the Principal Chiral model (PCM) (see e.g. \cite{kure, ogrewi}), given that both discrete integrable models and integrable field theories share the same algebraic content. A particularly interesting direction to pursue would be the study of integrable defects in connection with dynamical algebras. The first step towards this direction would be the identification of the associated defect matrices that is classification of generic representations of the dynamical algebras \cite{take}. A direct connection with 2D statistical modes such as the SOS/RSOS models would be then possible via the face-vertex transformation providing results of great physical as well as algebraic meaning. We hope to address these significant issues in the near future.

\end{document}